%% file: diffbom-ieee.tex
\documentclass[sigconf]{acmart}
\setcopyright{none}
\usepackage{subcaption}
\usepackage{multirow}
\usepackage{algorithmic}
\usepackage{graphicx}
\usepackage{textcomp}
\usepackage{xcolor}
\usepackage{svg}
\usepackage{xspace}
\usepackage{url}
\usepackage{listings}
\usepackage{enumitem}

\usepackage{tikz}

\newcommand{\darkfiles}[1]{\textsc{Darkfiles}\xspace}
\newcommand{\debian}[1]{200}

\definecolor{codegreen}{rgb}{0,0.6,0}
\definecolor{codegray}{rgb}{0.5,0.5,0.5}
\definecolor{codepurple}{rgb}{0.58,0,0.82}
\definecolor{backcolour}{rgb}{0.95,0.95,0.92}

\lstdefinestyle{newstyle}{
    backgroundcolor=\color{backcolour},   
    commentstyle=\color{codegreen},
    keywordstyle=\color{magenta},
    numberstyle=\tiny\color{codegray},
    stringstyle=\color{codepurple},
    basicstyle=\ttfamily\tiny,
    breakatwhitespace=false,         
    breaklines=true,                 
    captionpos=b,                    
    keepspaces=true,                 
    numbers=left,                    
    numbersep=5pt,                  
    showspaces=false,                
    showstringspaces=false,
    showtabs=false,                  
    tabsize=2
}

\lstdefinestyle{mystyle}{
    backgroundcolor=\color{backcolour},   
    commentstyle=\color{codegreen},
    keywordstyle=\color{magenta},
    numberstyle=\tiny\color{codegray},
    stringstyle=\color{codepurple},
    basicstyle=\ttfamily\footnotesize,
    breakatwhitespace=false,         
    breaklines=true,                 
    captionpos=b,                    
    keepspaces=true,                 
    numbers=right,                    
    numbersep=5pt,                  
    showspaces=false,                
    showstringspaces=false,
    showtabs=false,                  
    tabsize=2
}

\AtBeginDocument{%
  }
\begin{document}

\title{Software Dark Matter: Gazing at Uncharted Files to Navigate SBOM Integrations}

\author{Abhishek Reddypalle}
\affiliation{%
  \institution{Purdue University}
  \city{West Lafayette}
  \state{Indiana}
  \country{USA}
}
\email{areddypa@purdue.edu}

\author{Dennis Roellke}
\affiliation{%
  \institution{Bloomberg}
  \city{New York City}
  \state{New York}
  \country{USA}
}
\email{droellke@bloomberg.net}

\author{Santiago Torres-Arias}
\affiliation{%
  \institution{Purdue University}
  \city{West Lafayette}
  \state{Indiana}
  \country{USA}
}
\email{santiagotorres@purdue.edu}

\begin{abstract}
\input{sections/abstract}
\end{abstract}

\maketitle

\input{diffbom-body}


\bibliographystyle{plainnat}
\bibliography{diffbom}

\end{document}

%% file: sections/abstract.tex
Modern software supply chains have evolved into vast, heterogeneous networks where transparency — the granular understanding of all software components — is now a critical security requirement. While Software Bills of Materials (SBOMs) have emerged as the primary mechanism for this transparency, current industry practices rely on a metadata-centric paradigm that assumes an artifact is defined solely by its package manager declarations. We posit that this assumption is fundamentally flawed, creating a systemic visibility gap we define as Software Dark Matter (SDM). SDM represents the set of security-critical files present in an artifact's filesystem that are unaccounted for by its associated metadata. We implement a reference tool, DARKFILES, and use it to analyze four ecosystems of disjoint nature: DockerHub, Maven Central, plugin/extension marketplaces (Jenkins plugins and OpenVSX), and a real-world enterprise environment.

Our research makes the following contributions: we introduce a general-purpose metric for artifact fidelity calculating SDM as the ratio of untracked files per total file count. We introduce Packaging Lag, a phenomenon where official metadata remains out-of-date across multiple versions before catching up to an artifact's actual content. We demonstrate that SDM exposes vulnerable software invisible to SBOM-driven pipelines both by cross-referencing untracked packages against known CVE databases and through the direct discovery of three confirmed high-severity CVEs, showing that SDM is highly correlated with sensitive information including secrets and cryptographic keys.

%% file: diffbom-body.tex
\newcommand{\tool}{\textsc{Darkfiles}}
\newcommand{\osquery}{osquery}
\newcommand{\fleet}{FleetDM}

\section{Introduction}
\label{sec:introduction}
\input{sections/introduction}

\section{Background and Related Work}
\label{sec:background}
\input{sections/background}

\section{Software Dark Matter}
\label{sec:method}
\input{sections/method}

\subsection{Implementation}
\label{sec:implementation}
\input{sections/implementation}

\section{Analysis of Container Ecosystems}
\label{sec:results}
\input{sections/container}

\section{Analysis of Language Package Ecosystems}
\label{sec:maven}
\input{sections/maven}

\section{Analysis of Extensions Ecosystems}
\label{sec:extensions}
\input{sections/extensions}

\section{Analysis of Enterprise Systems}
\label{sec:enterprise}
\input{sections/enterprise}


\section{Eliminating Software Dark Matter}
\label{sec:discussion}
\input{sections/discussion2}

\section{Conclusions}
\label{sec:conclusions}
\input{sections/conclusion}



%% file: sections/introduction.tex
Modern software supply chains are large, complex networks that span multiple actors, organizations, jurisdictions, and ecosystems. 
Prior research has found that many projects have one hundred or more Open Source Software (OSS) components~\cite{ellison_evaluating_2010, ohm_backstabbers_2020, williams_2021_nodate, prana2021}.
This scale and heterogeneity have made software transparency, or the understanding of all software components and dependencies, a first-class software engineering and security concern~\cite{noauthor_synopsys_2021}.

For example, in December 2025, the react4shell software vulnerability affecting the react framework and next.js exposed a large number of modern web front-ends to remote code execution attacks~\cite{react4shell}. 
Industry response to this incident painfully demonstrated that even five years after the similar, infamous log4shell vulnerability~\cite{noauthor_nvd_nodate,noauthor_nvd_nodate_1,noauthor_cisa_2022} 
 there is no established automated solution to confidently determine - or reject - the presence of open source software dependencies within a software project~\cite{noauthor_apache_nodate,noauthor_review_2022, xia_icse_23, zahan_SP_23}. 

While various efforts in industry and academia have attempted to close this gap through numerous heuristics, there is no generalized model that allows a comprehensive exploration of this phenomenon.
This is primarily due to the fact that these studies focus on the effect (i.e., lack of transparency) rather than explore their causes.
Further, without a clear understanding of such causes, it is difficult to establish methods that address this lack of transparency.


In this paper, we hypothesize that existing solutions and operational pipelines fail for systemic reasons, in particular the informational divergence between a software manifest and its physical reality: Software Bill of Materials~(SBOMs)~\cite{ntia-sbom} have established themselves as a mechanism to provide much needed \emph{transparency} regarding the software stacks provided by vendors. 
The guiding principle behind SBOMs is that, when software vendors disclose a comprehensive view of the software components included in their products, software consumers would be able to take adequate action to minimize their attack surface. 
Thus, SBOMs are increasingly treated as a foundation for software supply-chain security~\cite{garcia2025landscape}. 
They are used to determine exposure during incident response (e.g., react4shell), to drive governance workflows, and importantly, justify risk claims about deployed artifacts. 


Unfortunately, the tools that populate SBOMs with information, often Software Composition Analysis (SCA) tools use an insufficient source of truth to establish software identity: These scanners generate post-hoc SBOMs that imply the physical presence of a component based on package manager databases and ecosystem-specific heuristics. 
While the approach of using tools that analyze package manager metadata has intuitive appeal, past research has suggested that they can introduce a degree of variability and inconsistency in their outcomes, resulting in dramatically different results~\cite{Imtiaz_2021, sbom_quality_2023, sca4container, wang_tosem_26, zhao_fse_2023}. 


However, this intuitive appeal is insufficient in real build and deployment workflows, for artifacts are assembled through heterogeneous origins not restricted to packages with metadata including scripts installing software outside package managers, broad copy/ unarchive steps include files without attribution, repackaging patterns (e.g., bundled dependencies and shaded artifacts) and post-deployment drift all introduce files after meta-data generation.

This work is the first to quantify this security gap using a metric that goes beyond listing the missing files; we introduce Software Dark Matter. 
SDM is the set of ~\emph{security critical} files actually present in an artifact’s filesystem namespace but not asserted from its SBOM. 
We implement this measurement by fundamentally shifting the source of truth behind our supply chain assertions: We leverage \emph{all} actual files of an artifact, and to the best of our knowledge we are the first to demonstrate a filtering strategy that amplifies the signals in this potentially noisy approach to draw the developers attention to important files only. 
This strategy has resulted in findings in popular open source projects, CVE-2025-32754 in Jenkins and CVE-2025-32111 in acme.sh, with further vulnerabilities currently under coordinated disclosure.
We note that SDM does not certify that an artifact is secure, instead it measures to what degree state-of-the-art software analysis (e.g., SBOM-driven) pipelines diverge from reality - what critical files they do not account for - and highlights them as suspicious. 
Without lightweight reachability analysis to prioritize SDM that is likely to influence behavior and surface security-relevant threats this approach would be little actionable for developers. 
We share \tool{} as reference implementation for this framework.



We use \tool{} to formally study the phenomenon of SDM by answering four leading research questions. 
Initially, research question (\textbf{RQ1}) tests our hypothesis that SDM is systemic to modern software by showing its prevalence across multiple representative ecosystems. Then, research question  (\textbf{RQ2}) affirms that SDM has security implications, and research question (\textbf{RQ3} and \textbf{RQ4}) provides academic insights about the origins of SDM - both causal and temporal.  

In conclusion, this paper presents the following contributions:
\begin{itemize}
    \item Analyzes a major root cause of insufficient software transparency despite years of attention and an abundance of regulations
    \item Identifies that real-world heterogeneous software to be oversimplified by regulators and vendors
    \item Formalizes Software Dark Matter a general-purpose metric for artifact fidelity, and discovers the phenomenon of \emph{Packaging Lag}
    \item Evaluates these measurements on a holistic software collection from disjoint real world sources, and 
    \item Provides tangible recommendations for improvement
\end{itemize}

   

The rest of the paper is organized as follows: Section~\ref{sec:background} provides background on software development practices, SBOM creation, and container technologies. Section~\ref{sec:method} defines the threat model, formally defines \textit{software dark matter} and presents theoretical ideas driving \tool{}. 
Section~\ref{sec:container}, Section~\ref{sec:maven}, Section~\ref{sec:extensions}  and Section~\ref{sec:enterprise} show our findings in the four ecosystems we study. 
Finally, in Section~\ref{sec:discussion} we present a strategies to eliminate SDM.

%% file: sections/background.tex

In industry, the state-of-the-art in software transparency is to use Software Composition Analysis (SCA) tools that programmatically aggregate dependency information as Software Bill of Materials (SBOM). 
While this approach is sound and complete for controlled experiments, related work illustrates its flaws in heterogeneous real-world software.
This section defines how SBOMs are generated, how SCA works, and the software ecosystems and practices in which these techniques are applied.

\subsection{SBOM Workflows in Supply-Chain Security}
SBOMs have emerged as a practical mechanism to enhance software security through transparency due to their ability to centralize all information about a software artifact in a machine-readable format.
Thus, they are integrated into vulnerability management, license checks, and policy decisions by means of direct references to claims, or by studying the inventory of software components and their relationships provided within.
In operational settings, SBOMs are treated as an input to automated security workflows, where they are (i) generated or collected after build, (ii) stored and distributed to downstream consumers, and (iii) ingested into vulnerability management and policy engines for remediation and compliance reporting.
NIST frames SBOMs as a means to increase transparency and accelerate identification and remediation of vulnerabilities in supply chains, emphasizing their utility as an operational artifact rather than a purely documentary one~\cite{maratos2025supply}.

As SBOMs become embedded into Continuous Integration and Continuous Delivery(i.e CI/CD), security teams increasingly rely on them for evidence regarding claims about what is present in deployed software, assess risk, and respond to incidents when new vulnerabilities emerge.
Industry and policy developments have further accelerated adoption. 
In the United States, the U.S.\ Executive Order EO14028 ecosystem established SBOMs as a procurement-facing control and White House guidance has required or encouraged agencies to obtain SBOMs for certain software procurements~\cite{eo14028, omb23-16, omb22-18}.
Adoption surveys also suggest that SBOM generation and consumption are becoming increasingly common in practice~\cite{openssf-sbom, EUhasone, sonatype_sbom_survey, stalnaker_icse_24}.

Similarly, in the European Union, EU's Cyber Resilience Act (CRA) introduces lifecycle obligations around vulnerability handling and reporting, and specifies a phased applicability timeline (entered into force on 10 December 2024, reporting obligations apply from 11 September 2026, main obligations apply from 11 December 2027)~\cite{eu_cra_timeline}.
Such regulatory pressures further heighten the need for measurable SBOM accuracy.
Without a defensible accuracy baseline, policy requirements are difficult to enforce in practice. 
SBOM producers can plausibly attribute missing components to tool limitations or ecosystem-specific coverage gaps, shifting accountability to the capabilities of SBOM generators instead of the producing pipeline~\cite{melara2023, sbom_quality_2023}.

However, as SBOMs play a crucial role in security decision-making, it is important to understand \emph{to what extent do SBOM-driven pipelines faithfully represent the artifacts they are used to secure?}

\subsection{Software Composition Analysis}
\label{sec:background:sca}

Software Composition Analysis (SCA) tools mine dependency information from software artifacts and aggregate dependency information, licensing data, and known vulnerabilities into a standardized format called a Software Bill of Materials (SBOM).
US federal bodies have shaped SBOM expectations at multiple levels:
CISA distinguishes multiple lifecycle-aligned SBOM types---including \emph{design}, \emph{source}, \emph{build}, \emph{analyze}, and \emph{runtime/deployed} SBOMs---that differ in visibility and intended use~\cite{cisa_sbom_types}, while NTIA's 'minimum elements' guidance, issued under Executive Order 14028~\cite{eo14028}, provides a baseline set of required data fields but does not establish what makes an SBOM accurate~\cite{ntia-sbom}.
In practice, most SBOMs are generated post-build as \emph{analyze} SBOMs using package metadata and ecosystem heuristics~\cite{Balliu_2023,halbritter2024}.
These limitations motivate a shift from schema- and completeness-oriented assessments toward artifact-grounded measures of SBOM accuracy for security decision-making to better achieve an SBOM's stated mission of a comprehensive risk assessment clearinghouses.

As SBOMs are increasingly positioned as evidence supporting these obligations, pipelines need a way to quantify what SBOMs \emph{do not} capture, not only what they claim to include.
Our approach is different in that, rather than inferring provenance from metadata alone, we compare the actual filesystem of the deployed artifact to SBOM claims, surfacing all files whose provenance is unclear or absent, regardless of how or why they were introduced.

Comparative studies consistently find large disagreement among SCA tools run on the same project, driven by differences in detection heuristics and advisory sources. 
Imtiaz et al.\ report counts ranging from 17 to 332 vulnerable dependencies across nine tools on the same Maven/npm application and recommend against relying on any single scanner~\cite{Imtiaz_2021, dann_tse_2022}. In containers, Churakova \& Ekstedt measure low consistency among VEX-enabled container scanners, indicating low maturity of the space~\cite{churakova2025}. Tooling reports like the Container SBOM Clarity Project similarly document inconsistent package identification and licensing in container SBOMs~\cite{container_sbom_clarity2024}. Broad surveys also note substantial variability and reproducibility challenges in SBOM outputs across tools and versions~\cite{sharma2025,rabbi2024,yu2024,sbom_quality_2023,eric2024,omar2021,halbritter2024}.

Even when tools agree on approach, output can vary across formats and versions. Balliu et al.\ deep-dive Java SBOM producers and highlight reproducibility/accuracy gaps, reinforcing the need for maturity before SBOMs can serve as reliable ground truth~\cite{Balliu_2023, wang_tosem_26}. Multiple formats (SPDX, CycloneDX) and identifier schemes (e.g., SWID, PURLs) introduce cross-ecosystem matching problems when not uniformly adopted~\cite{purl_specification, ozkan2025supplychaininsecuritylack}. Melara and Torres-Arias argue that many challenges stem from the lack of a common language and shared understanding around supply-chain metadata~\cite{melara2023}.

Large-scale measurements show that many popular images contain manually installed software that common SBOM generators overlook. Kawaguchi et al.\ analyze 3{,}500{+} DockerHub images and find that 51\% include one or more manually installed packages; between 30–70\% of these installations are missed by prominent SBOM tools, and 22.7\% of actively executed packages go unidentified—including some with known CVEs~\cite{kawaguchi2024,prana2021,williams2025, eric2024, bufalino_2025}. Dockerfile-aware/static systems like DAVS reason from Dockerfiles to locate potentially vulnerable files but assume access to build artifacts and focus on targeted vulnerability inference rather than SBOM validation~\cite{doan2022}. 
In contrast, this work is proactive and CVE-agnostic, and thus complements the approaches listed above. 
Rather than reacting to known issues via a vulnerability database, it surfaces integrity and provenance gaps that can be acted upon even before any CVE or 0-day is disclosed.


\subsection{Heterogeneous Real-World Software}
\label{sec:background:enterprise}
Recent work shows how metadata-centric pipelines under-report or mis-attribute components when confronted with packaging indirection.
Dietrich et al.\ show that hidden dependencies introduced via cloning or shading (e.g., class relocation, name mangling) escape mainstream SCA and surface as vulnerable ``clones'' missed by tools~\cite{dietrich2024}.
Adversarial analyses in the Maven ecosystem demonstrate that POM indirections, dependency management inheritance, and shaded/uber-JAR repackaging can confuse or mislead meta-data based SCA (e.g., Dependabot, Snyk, Grype, OSV-Scanner, OWASP Dependency-Check), yielding false-negatives and other failure modes~\cite{ivanova2025adversarial}.
Recent works also highlight parser confusion attacks, where inconsistencies among toolchain parsers are exploited to conceal components or inject ambiguity into SBOMs~\cite{yu2024}.

\begin{figure}[t]
\centering
\includegraphics[width=.8\linewidth]{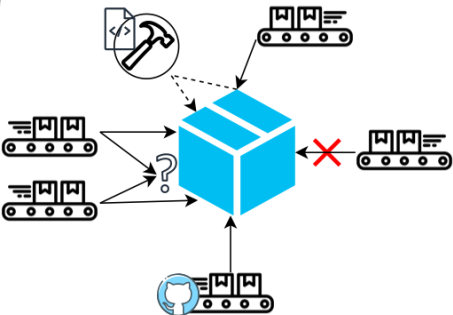}
\caption{Real world software composition with (A)~build-context bleed-in adding unintended files (B)~provenance loss due to repackaging practices stripping attributes from the code (C)~divergence between used manifest and declared manifest (D)~post-build drift added by runtime initialization}
\label{fig:ecosystems}
\end{figure}

These challenges extend beyond third-party analyzers: ecosystem and registry metadata can itself be missing, inconsistent, or incorrect, leading to provenance drift between source and published artifacts.
For example, PyRadar and LastPyMile (Python) and AROMA (Java) show that even first-class metadata (e.g., the one used by a package manager) can be incomplete or wrong, which can undermine provenance and cause mismatches between source-level intent and released artifacts~\cite{pyradar,lastpymile,AROMA, goswami_icsme_2020, mujahid_ase_22}.
Finally, several studies document the lack of integrity protection and poor cross-ecosystem meta-data hygiene in current SBOM solutions, compounding the risk of incomplete or misleading inventories~\cite{ozkan2025supplychaininsecuritylack, 10179320}.


Further, production grade enterprise systems are compositions of applications stacked vertically and horizontally and in practice maintained by independently operating teams.
One team may use multiple pipelines, and each pipeline may be used by multiple teams. 
Multiple pipelines may contribute to the same layer of abstraction, or one pipeline may serve multiple layers. 
In practice, these pipelines are implemented through manifest files (e.g., build specs, deploy descriptors, policy as code).
The prior discussion about which SBOM types to employ must therefore scale from \emph{one} pipeline to \emph{many} pipelines to obtain system-wide coverage. 
However this is difficult in practice, existing tooling tends to center on application workflows and ecosystems with package managers, leaving gaps when artifacts are assembled by heterogeneous pipelines and ad-hoc steps.
Container images(represented as layered OCI manifest or commonly Dockerfiles) exhibit the \emph{same multi-party, multi-pipeline structure}~\cite{oci}. 
Each Dockerfile instruction materializes as a filesystem layer, much like a CI job stage, so the final image reflects the union of decisions made across pipelines.
This is precisely where discrepancies arise: images routinely ship components installed outside package managers (manual copies, \texttt{curl|bash} installers, compiled-from-source), or vendored/shaded artifacts carried forward from upstream layers. 
Finding SDM in modern systems inherently requires recursive, context-aware analysis, including across container layers, deeply nested infrastructure-as-code pipelines, or packaging formats like shaded JARs~\cite{maven_shade_plugin,diffoscope}. 
Our approach generalizes to all these scenarios, rather than relying on a single manifest or ecosystem’s perspective.

%% file: sections/method.tex
In this section we formalize the problem by defining a threat model, a definition of software dark matter, and the driving research questions for our study.

\subsection{Threat Model}
A vast body of work in software supply chain security focuses on a \emph{reactive} threat centered around vulnerability propagation by studying how known software vulnerabilities, e.g. CVEs, or known malicious packages~\cite{decarli-combosquat, zimmermann_2019} propagate on to different systems and how to manages ensure proper incident response~\cite{ohm_backstabbers_2020, ladisa_taxonomy_2022}. 
Instead, our work lies on the field of supply chain hardening, that is to \emph{proactively} eliminate supply chain vulnerabilities added by natural occurrence. 
The SDM framework does no expect prior knowledge of vulnerabilities or malicious packages and it is not a defense against above mentioned distribution vectors~\cite{okafor_scored_22}. 
Within this hardening space, we focus on unknown transparency gaps that introduce vulnerabilties. 

\paragraph{Attacker Goals \& Motivation}
Explicitly, we assume an attacker, or negligent user seeks to compromise a software artifact in the software supply chain by introducing underhanded changes that are not reported by SCA tools. 
Given the transitive nature of software supply chains, attacks are not necessarily focused on a specific victim but may exploit a given CVE on any system that's subject to it to maximize their reach.

By operating within this space, adversaries can:

\begin{itemize}
    \item Reason about vulnerabilities: Ideally the adversary finds a CVE that the system owner is not aware of and has not mitigated.
    \item Backdoor Credentials: Extract secrets unintentionally inserted into images, such as leaked GitHub tokens or pre-generated SSH keys, which are often untracked by package managers~\cite{markus2023, meli_ndss_2019}.
    \item Evade Detection: Hide malicious components, vulnerable libraries (e.g., vulnerable copies of Log4j), or persistence mechanisms 
\end{itemize}

\paragraph{Attacker Strategy}

Attackers can exploit the fact that SBOMs are often incomplete because the underlying metadata is itself incomplete.
As explored in the paper below, we identify four recurring patterns through which this gap manifests, each observed across multiple ecosystems.
We exemplify these patterns as follows:

\begin{enumerate}[label=\alph*]
\item \textbf{Build-context bleed-in.}
Broad copy or packaging steps pull unintended files into the artifact. In containers, the \texttt{acme.sh} image's \texttt{COPY ./ .} captured a \texttt{.git/config} containing a GitHub token with push privileges (CVE-2025-32111). The same pattern appears in VS~Code extensions that ship \texttt{.git/}, \texttt{.env}, and \texttt{.ssh/} directories (\ref{sec:extensions}).
 
\item \textbf{Provenance loss.}
Repackaging strips dependency attribution, leaving embedded code untracked. In Maven, shaded uber-JARs relocate and rename classes, erasing the original package namespace and POM provenance trail (\ref{sec:maven}). In VS~Code extensions, JavaScript bundlers concatenate all dependencies into a single minified file, destroying per-package identity entirely (\ref{sec:extensions}).
 
\item \textbf{Metadata divergence.}
A bundled dependency's actual version disagrees with or is absent from the artifact's declared metadata. Across our Jenkins plugin corpus, 57 plugins ship a JAR in \texttt{WEB-INF/lib/} at a version that differs from what the POM declares(the SBOM reports a safe version while the physically bundled JAR is vulnerable (\ref{sec:extensions})). In containers, our longitudinal analysis reveals untracked files in one release are only incorporated into official packages several versions later (\ref{sec:container}).
 
\item \textbf{Post-build drift.}
Build-time side effects or runtime initialization produce files that persist without metadata coverage. The Jenkins \texttt{ssh-agent} image bakes SSH host keys generated during \texttt{apt install openssh-server} into the published image rather than deferring generation to runtime, enabling impersonation across container instances (CVE-2025-32754, CVE-2025-32755). In enterprise environments, IaC provisioning retrieves binaries from remote hosts outside the package manager, producing files whose provenance is captured by no metadata channel (\ref{sec:enterprise}).
\end{enumerate}

These patterns share a common structure: files are physically~\footnote{by physical, we refer to the fact that it is tangibly contained in the artifact, rather than exclusively reported} present in the artifact yet absent from its declared metadata.
However, they arise from distinct pipeline stages and packaging conventions, and they recur across technically disjoint ecosystems.
Metadata-centric SCA tools cannot surface them by design, because these tools analyze declared state rather than physical contents.
Ignoring this gap leads to a false sense of security: security teams may certify an artifact as vulnerability-free because their scanners only see tracked metadata, while high-severity risks(including those with CVSS scores up to 9.1) in the untracked files. 
The \tool{} framework addresses this by alleviating an SBOM's false-negatives, improving completeness rather than soundness.




\subsection{Definition}
\begin{figure*}[htbp] 
\centering
\includegraphics[width=.85\textwidth]{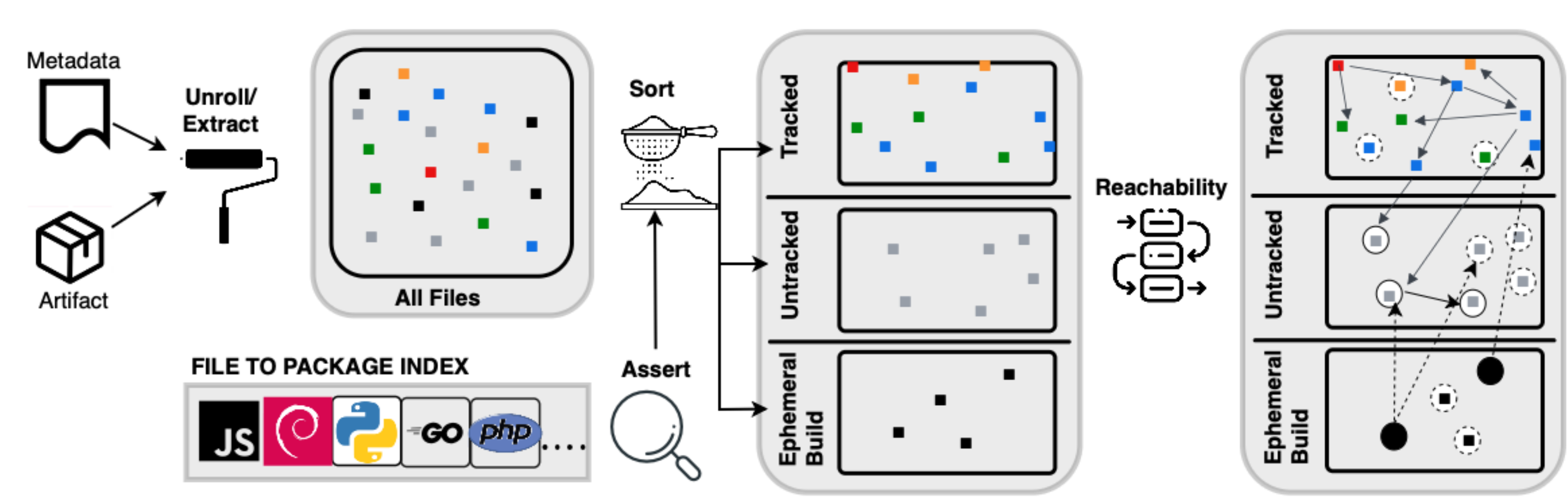}
\caption{\tool{} analysis flow. The first passes obtain an exhaustive list of artifacts, which are then sorted into categories. Afterwards, a reachablity and influence analysis prioritizes SDM that may alter the behavior of a target artifact.}
\label{fig:filter}
\end{figure*}

To measure this phenomenon, we propose a general-purpose software quality metric calculated at the file level, quantifying the reach of metadata from the underlying package manager as the percentage of files not tracked by it. 
We first define the set of \emph{untracked files} as $\mathit{Files}_{\mathit{actual}} \setminus \mathit{Files}_{\mathit{asserted}}$: files physically present in the artifact's filesystem namespace but not asserted by its associated metadata. 
These files capture the raw provenance gap. However not all untracked files are equally concerning. 
We therefore define \emph{Software Dark Matter}, SDM, as the security-relevant subset that survives a static reachability and influence filter~$\Phi$: files that are transitively reachable from an artifact's entrypoints, or that exhibit sensitivity signals such as cryptographic key material or credentials.

\begin{equation}
\mathcal{SDM} = \Phi(\mathit{Files}_{\mathit{actual}} \setminus \mathit{Files}_{\mathit{asserted}})
\end{equation}

\noindent where $\Phi$ is the static reachability filter defined in Section~\ref{sec:impl:reachability}. 
Throughout this paper, when we report prevalence figures, we state explicitly whether a given measurement refers to the broader set of untracked files or to the reachability-filtered SDM.

\begin{table*}[]
\centering
\begin{tabular}{r|llll}
    \toprule
 &
  \multicolumn{1}{c}{\textbf{RQ1: Prevalence}} &
  \multicolumn{1}{c}{\textbf{RQ2: Security}} &
  \multicolumn{1}{c}{\textbf{RQ3: Origin}} &
  \multicolumn{1}{c}{\textbf{RQ4: Half-Time}} \\ \hline
\textbf{Container} &
  \begin{tabular}[c]{@{}l@{}}Power-Law shows \\ fewer containers \\ have high SDM, \\ but still common\end{tabular} &
  \begin{tabular}[c]{@{}l@{}}3 zero days reported,\\ associated CVEs have\\ High and Critical \\ CVSS scores\end{tabular} &
  \begin{tabular}[c]{@{}l@{}}File system layers \\ like ADD and COPY \\ introduce SDM\end{tabular} &
  \begin{tabular}[c]{@{}l@{}}SDM lags latest versions \\ before it's covered \\ by meta-data,\\ i.e. Packaging lag\end{tabular} \\ \hline
\textbf{Package} &
  \begin{tabular}[c]{@{}l@{}}Bi-Modal shows\\ native packages either \\ no SDM or high SDM\end{tabular} &
  \begin{tabular}[c]{@{}l@{}}100s of CVEs undetected\\ by SCA tools, incl. \\ widely used projects\end{tabular} &
  \begin{tabular}[c]{@{}l@{}}Nesting/ Shading \\ introduce SDM\end{tabular} &
  \begin{tabular}[c]{@{}l@{}}SDM does not change\\ over time, supporting that \\ it's intentional/ systemic\end{tabular} \\ \hline
\textbf{Plugins} &
  \begin{tabular}[c]{@{}l@{}}Production grade \\ obfuscation methods\\ skew the analysis; \\ source maps restore \\as observability side channel \end{tabular} &
  \begin{tabular}[c]{@{}l@{}}about 1000 of CVEs \\ are SBOM-invisble ; \\ credential leak\end{tabular} &
  \begin{tabular}[c]{@{}l@{}}Bundling/Vendoring \\ dependencies\\ introduces SDM\end{tabular} &
  Not representative \\ \hline
\textbf{Enterprise} &
  {[}redacted{]} &
  \begin{tabular}[c]{@{}l@{}}Detected unused but \\ executable turnscripts\end{tabular} &
  \begin{tabular}[c]{@{}l@{}}IT automation/ \\ Config management\\ introduces SDM\end{tabular} &
  {[}redacted{]} \\
  \bottomrule
\end{tabular}
\caption{Summary of findings from Research Questions per Ecosystem}
\label{tab:findings}
\end{table*}

\subsection{Research Questions}
To apply our measurement of Software Dark Matter, we examine ecosystems that excercise different parts of the modern software supply chain:
(i) Container images from DockerHub, which serve as foundational base environments and are integrated into automated build workflows 
(ii) Maven Central artifacts, which are the primary distribution mechanism for Java libraries and undergo repackaging practices as part of their release pipelines
(iii) Extension ecosystems such as Jenkins plugins and VS Code extensions, which are loaded by host applications at runtime and follow their own packaging and review processes independent of the host's dependency graph
(iv) A centrally managed enterprise Linux environment, where long-lived systems evolve through operational updates, provisioning workflows, and ongoing reconfiguration.
Together, these settings capture image-based, library-based, host-mediated, and configuration-driven modes of software delivery and allow us to evaluate how SDM manifests across heterogeneous pipelines.
Our approach using \tool{} generalizes to all four ecosystems, as shown in Figure~\ref{fig:filter}.
We then utilize \tool{} to investigate the following questions:

\begin{itemize}
    \item \textbf{RQ1 Is Software Dark Matter a prevalent phenomenon?}
We quantify the extent to which the SBOM-asserted state diverges from the observed contents of an artifact, measuring the set of files present in the artifact but not accounted for by its metadata, to motivate the need for improved transparency and tooling and to validate the proper integration and usage of those tools.

    \item \textbf{RQ2 What are the security implications of Software Dark Matter?} 
We apply the reachability and influence filter to the untracked files surfaced by RQ1 to obtain the SDM subset, and explore its potential security impact. Establishing whether SDM harbors security-relevant exposures enables stakeholders to recognize why the problem needs to be addressed and then to address these hidden risks proactively, ultimately to build better trust and compliance within software ecosystems.

    \item \textbf{RQ3 Where does Software Dark Matter originate? What actions cause it?}
We identify the specific development, build, and distribution practices that produce SDM in each ecosystem, including which build stages introduce it, how it evolves, and which packaging conventions structurally obscure provenance.

    \item \textbf{RQ4 How does a project's Software Dark Matter change over time?}
    We identify trends in dark matter prevalence in artifacts over time, to gain insight on whether it behaves similarly to other software metrics which often increase or decrease with a project's longevity.

\end{itemize}

Our work explicitly prioritizes identifying, measuring, and highlighting the phenomenon of Software Dark Matter rather than providing a comprehensive solution. 
We focus on demonstrating the existence and impact of dark matter to encourage future research and development aimed at addressing these critical gaps in current SBOM generation and management tooling.

%% file: sections/implementation.tex
We implemented \tool{} as an open-source, modular analysis tool that identifies untracked files and applies the reachability filter~$\Phi$  to surface SDM. The analysis workflow proceeds through the passes illustrated in Figure~\ref{fig:filter}.

\paragraph{Unroll and Extraction}
The first pass captures a complete, structured view of the target filesystem. 
For container images, \tool{} pulls (i.e, downloads) and iterates over an image's Open Container Initiative (i.e, OCI) layers to extract the files at each intermediate step. 
The layered filesystem architecture preserves differential changes(added, modified, or deleted artifacts) which enables per-layer analysis and correlation of untracked files with specific Dockerfile instructions. 
The same recursive-extraction approach generalizes to JAR archives (resolving nested JARs and transitive dependencies), extension packages, and multi-source manifest aggregation in enterprise environments similarly to Lamb et al~\cite{lamb2021reproducible}.

\paragraph{Large Scale File to Package Index}
\label{sec:impl:known}
A raw filesystem contains many routine artifacts like caches, logs, packaging-convention paths that must be distinguished from genuinely untracked files. 
\tool{} builds a compound \emph{known-files} filter through three layers.
A general filtering pass first removes transient noise (temporary files, logs, caches). 
Then, a precomputed file-to-package pass identifies files recognized by OS-level package managers: for Debian-based systems, our \texttt{file2pkg} module queries distribution repositories offline and generates a comprehensive SQL-based mapping database, currently spanning 5~major Debian and 6~Ubuntu distributions (${\sim}$100\,GB of indexed data), providing deterministic provenance lookups far faster than per-file \texttt{apt-file} queries. 
Finally, configurable regex-based passes capture files declared or routinely produced by language-level tooling (e.g., \texttt{site-packages/} for Python, \texttt{node\_modules/} for JavaScript, \texttt{target/} and JAR layouts for Java. Paths matching these patterns are marked as known with provenance annotated as \emph{language-rule}, aligning with the outputs that SBOM generators derive from manifest files such as \texttt{requirements.txt}, \texttt{package.json}, and \texttt{pom.xml}.
This classification is not a security boundary, an attacker who places a malicious file inside a tracked path (e.g., a backdoored package in \texttt{node\_modules/}) does not evade detection, because that file falls within the scope of conventional SCA tools, which already index and scan these locations.
The filtering step is a \emph{scoping} decision for \tool{} to measure the gap that existing tooling leaves, not the space it already covers.

\paragraph{Assertion and Surfacing}
\label{sec:impl:sorting}
After filtering, \tool{} produces a \emph{KnownFiles} set where each path is linked to zero or more claimants (OS packages or language rules) and tagged with the pass that established the claim. 
Files are then sorted into two bins: \emph{tracked} (claimed by at least one pass) and \emph{untracked} (no claimant). 
For the reachability analysis: tracked files serve as traversal nodes in the call graph, while untracked files are the candidates that the analysis seeks to flag.

\paragraph{Reachability and Influence}
\label{sec:impl:reachability}

Not all untracked files are equally concerning, a stray log template poses less risk than an untracked shared library loaded at startup.
To distinguish the two, \tool{} performs a static reachability analysis that identifies which untracked files can influence runtime behavior.
The process begins with entrypoint discovery: identifying primary scripts, binaries, service units, or container entry commands by inspecting configuration files, manifests, and build metadata. 
Starting from each entrypoint, \tool{} traverses reachable files using a conservative static call/reference graph whose edges include:

\begin{itemize}
  \item \textbf{Interpreter and import chains:} Following \texttt{\#!} shebang lines, language-level imports (e.g., Python, Node.js), shell \texttt{source} statements, and similar inclusion mechanisms.
  \item \textbf{Binary linkage:} Examining dynamic dependencies for ELF, PE, and Mach-O binaries, including interpreter paths, RPATH, \texttt{LD\_LIBRARY\_PATH} hints, and linked libraries~\cite{man7-ldso8}.
  \item \textbf{Configuration references:} Scanning known config formats, entrypoint definitions, and manifest files for direct or indirect file-path references.
\end{itemize}

When this traversal crosses from a tracked node into an untracked file, that file is classified as SDM and prioritized for review.
The distinction matters since tracked files along the path are already visible to SCA tools and do not require additional attention, whereas the untracked files they reach represent blind spots that no metadata-driven pipeline would surface.
For layered artifacts such as containers, \tool{} additionally performs a \emph{deep-darkfiles} analysis: it tracks files introduced in one layer and removed in a subsequent layer (e.g., via whiteout), surfacing build-time artifacts that persist in intermediate layers despite apparent deletion.

\tool{} is deliberately simple in its design. 
This simplicity allows it to generalize to various disjoint ecosystems with only policy-level configuration changes.
Our evaluation spans thousands of artifacts across four ecosystems, a scale heavier-weight techniques would struggle with. 
Even with the design choice, the utility is evident given that we surfaced zero-day and thousands of previously invisible vulnerabilities. 

\paragraph{Limitations}
\label{sec:impl:limitations}
The filtering passes in \tool{} are configurable, allowing organizations to add custom exclusions tailored to their environments (see Section~\ref{sec:enterprise}, where we apply this). Because the filter configuration is a policy parameter, prevalence figures reported throughout this paper represent a conservative, policy-driven lower bound rather than a definitive count. Tightening or loosening filters shifts the reported SDM percentage, so readers should interpret our numbers as a baseline under the specific filter set we describe.

Our static reachability component favors precision over recall. It primarily resolves absolute paths and does not model argument-dependent \texttt{exec} calls or complex shell dynamics. Reachability tags should therefore be read as a conservative lower bound on the true reachable set. We prefer to miss a subset of dynamically reachable files rather than over-claim reachability.

\begin{figure*}
\begin{minipage}[b]{0.3\textwidth}
\begin{lstlisting}[caption={List of files from unrolled container layers}, label={lst:code1}]

# Layer:2|Size:132.5 MB|

Command:    RUN apt-get install ...

    /etc/ca-certificates.conf
    /etc/gitconfig
    /etc/group
    /etc/gshadow
    /etc/hosts.allow
    /etc/ssh/sshd_config
    /usr/bin/nc.traditional
    /usr/sbin/sysctl
    /var/lib/dpkg/alternatives/nc
    ...
\end{lstlisting}
\end{minipage}%
\hfill 
\begin{minipage}[b]{0.3\textwidth}
\begin{lstlisting}[caption={Callgraph}, label={lst:code2}]
{
    "image": "jenkins/ssh-agent:6.11.1",
    "summary": {
        "entrypoint_main": "setup-sshd",
        "roots": ["/usr/local/bin/setup-sshd"],
        "reachable_nodes": 6,
        ...
    },
    "callgraph": {
        "/usr/local/bin/setup-sshd": [
            "/usr/sbin/sshd"
        ],
        "/usr/sbin/sshd": [
            "/lib64/ld-linux-x86-64.so.2",
            "/usr/lib/openssh/ssh-sk-helper"
        ...
}
\end{lstlisting}
\end{minipage}%
\hfill 
\begin{minipage}[b]{0.3\textwidth}
\begin{lstlisting}[caption={Darkfiles Deep-Report}, label={lst:code3}]
{
"summary": {
    "layers_dir": "...",
    "num_layers": 7,
    "introduced_candidates": 891,
    "deep_darkfiles": 266
},
    "items": [
        {
        "path": "/install_acme.sh/.git/config",
        "introduced_at_layer": 2,
        "removed_at_layer": 3,
        "removed_reason": "whiteout",
        "introduced_by": "...",
        "sensitivity_hint": [ "git_config" ]
        }
}
\end{lstlisting}
\end{minipage}
\end{figure*}

%% file: sections/container.tex
\label{sec:container}
Container images assemble software through heterogeneous pipelines like package-manager installs, ad-hoc downloads, multi-stage copies, and post-build configuration layered into a single deployable artifact. Because each mechanism can introduce files independently, the final image often contains components that no single metadata source fully describes. This makes containers a natural first target for measuring the gap between what metadata asserts and what the filesystem contains.\cite{liu_ESORICS_2020, shu_CODASPY_2017}

The SCA tools most commonly applied to containers inherit the meta-data centric assumption discussed in Section~\ref{sec:background}: they identify packages by querying distribution databases and language-specific manifest paths, so any file installed outside those channels is invisible to the resulting SBOM. The analysis below uses \tool{} to quantify, triage, and trace the origins of that invisible remainder.

\paragraph{Data Collection and Methods}
\label{sec:eval:methods}
We analyzed Debian- and Ubuntu-based images from the top 5,000 most-downloaded containers on DockerHub~\cite{ecosyste_ms}. This collection spans minimal base images to complex multi-service applications and provides a representative cross-section of real-world container adoption. We focus on Debian-family distributions because of their widespread usage in enterprise and open-source projects (approximately 50\% of the top 5,000 images) and the availability of high-quality package metadata from which we build the file-to-package mapping infrastructure. Distribution classification relies on the \texttt{os-release} file in each image's base layer~\cite{os-release}.

Once retrieved, we apply the \tool{} workflow of Section~\ref{sec:implementation}: each image's layered filesystem is unrolled with per-layer context preserved, then passed through the filtering pipeline. For OS packages, file-to-package provenance is derived \emph{deterministically} from official distribution mirrors, yielding stable, reproducible mappings for a given release. For language ecosystems, we apply per-language path rules that mirror how {SBOM generators recognize package installations. Table~\ref{tab:language-patterns} lists the exact regexes used for each ecosystem (those accounting for ${>}1\%$ of file removals) and the intent of each rule, allowing adaptation to other environments or SBOM pipelines.

\begin{table}[h]
  \centering
  \scriptsize
  \setlength{\tabcolsep}{2pt}
  \renewcommand{\arraystretch}{1.12}
  \begin{tabular}{@{}p{1.2cm}p{5.4cm}r@{}}
    \toprule
    \textbf{Language} & \textbf{Pattern} & \textbf{Hit-Rate} \% \\
    \midrule
    Javascript\cite{npm_folders} & \texttt{\string^.*node\_modules/.*\string} & \textbf{57.1909} \\
    Python\cite{python_site} & \texttt{\string^/usr/(local/)?lib/python\string\d+\string.\string\d+/.*\string} & \textbf{22.5072} \\
    Go\cite{go_goroot} & \texttt{\string^/usr/local/go/.*\string} & \textbf{10.4461} \\
    Node\cite{npm_folders} & \texttt{\string^/(root|home/[\string^/]+)/.npm/.*\string} & 1.1945 \\
    Python\cite{pep370} & \texttt{\string^/(root|home/[\string^/]+)/.local/lib/python\string\d+\string.\string\d+/.*\string} & 1.0190 \\
    Ruby\cite{rubygems_paths} & \texttt{\string^/usr/local/lib/ruby/gems/.*\string} & 1.0417 \\
    \bottomrule
  \end{tabular}
  \caption{\textbf{Language specification based ignore-patterns}}
  \label{tab:language-patterns}
\end{table}

\subsection{RQ1: Prevalence of Untracked Files}
\label{sec:eval:prev}
Figure~\ref{fig:docker-hub-images} shows the distribution of untracked-file percentages across the corpus. The distribution is wide: approximately 22\% of images have less than 5\% untracked files, yet no image has zero. Treating each container equally, the mean untracked-file percentage is 30\% and the median is 21\%, weighting by total file count raises the mean to 37\%. We exclude the first layer (the \texttt{FROM} base image or a \texttt{COPY --from=} multi-stage import) from these calculations because any untracked files introduced there are inherited by all downstream layers and thus attributable to the base-image supplier rather than the image author.

\begin{figure}[t]
    \centering
	\includegraphics[trim=10 20 10 10, width=.8\linewidth]{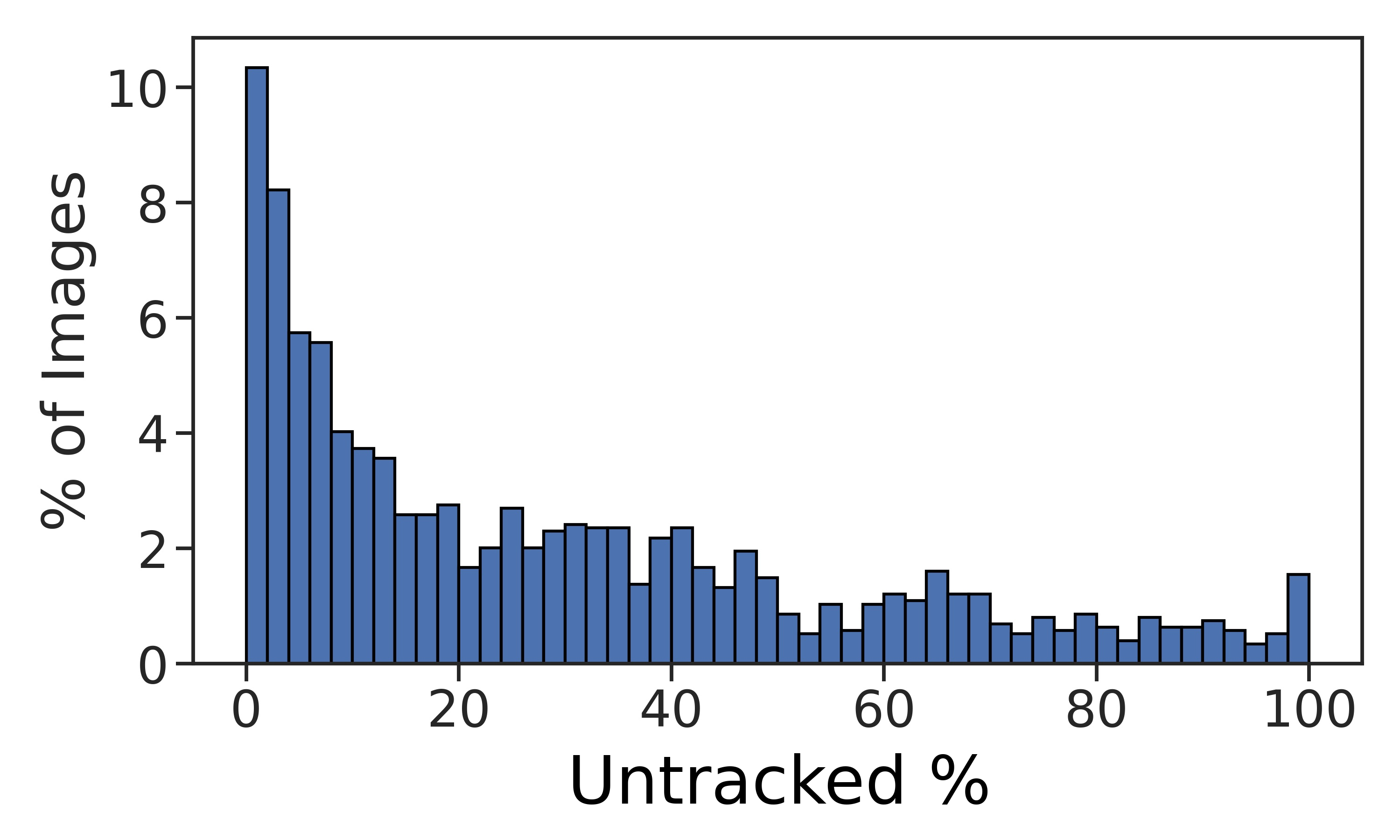}
	\caption{Untracked files in Popular DockerHub Images}
	\label{fig:docker-hub-images}
\end{figure}
~\\
\begin{table*}[]
\centering
\footnotesize
\setlength{\tabcolsep}{4pt}
\renewcommand{\arraystretch}{1.15}
\begin{tabular}{p{0.24\linewidth} p{0.1\linewidth} p{0.58\linewidth}}
\hline
\textbf{File} & \textbf{Images} & \textbf{Description \& security impact} \\
\hline
\texttt{/opt/bitnami/scripts/liblog.sh} & 63 & Bitnami logging helper; tampering can suppress/audit-bypass logs or inject startup commands. \\
\texttt{/usr/bin/with-contenv} & 30 & s6-overlay env injector; can alter PATH and redirect exec before the main service. \\
\texttt{/etc/supervisord.conf} & 28 & Supervisor config; can start extra daemons or change args/logging. \\
\texttt{/etc/ssh/ssh\_host\_rsa\_key} & 7 & SSH host private key; baked/reused keys enable impersonation/MITM across deployments. \\
\texttt{/usr/sbin/enable\_insecure\_key} & 13 & “Insecure key” toggle; reachable path to downgrade trust/auth (a security-off switch). \\
\hline
\end{tabular}
\caption{Most frequently seen SDM in dockerhub and why it matters.}
\label{tab:rq4-referenced-artifacts-halfcol}
\end{table*}

\subsection{RQ2: Security Impact}
\label{sec:eval:security}
Not all files are equally concerning. 
Through \tool{}'s entrypoint-rooted reachability analysis (Section~\ref{sec:impl:reachability}) we can narrow the triage set from the full untracked-file population of RQ1 to the SDM subset(files that are likely to influence runtime behavior). 
Across the corpus, this reachability filter reduces the triage space to \textit{approximately 1\% of files} on average, making manual review practical.

Table~\ref{tab:rq4-referenced-artifacts-halfcol} lists SDM files that recur across multiple images and that the reachability pass identifies as security-critical, including security-sensitive configurations and key material (e.g., \texttt{sshd\_config}, \texttt{ssh\_host\_rsa\_key}). 
These artifacts control environment initialization and service startup. 
If left untracked, they become blind spots for SBOM-based review despite being \emph{directly} referenced from executable code. 
We further enrich the reachability triage with secret-detection signals informed by prior large-scale studies showing that container images frequently embed secrets and credentials~\cite{hequan2024, markus2023}. 
Together, the reachability, provenance, and sensitivity signals discovered the following vulnerabilities.

\noindent\textbf{Static SSH Host Keys in Jenkins Container Images (CVE-2025-32754, CVE-2025-32755):}
Our analysis detected a vulnerability in the widely used Jenkins \texttt{ssh-agent} image. Jenkins, a leading open-source automation server with extensive enterprise usage for continuous integration and delivery pipelines, leverages the \texttt{ssh-agent} image to delegate workloads securely from controllers to agents. 
However, the \texttt{ssh-agent} image included pre-generated SSH host keys created during a build step that executed \texttt{apt install openssh-server}. 
These keys are intended to be generated at runtime to ensure uniqueness per deployment. 
However, since they were included in the image during build, identical keys are reused across every container instantiated from it. 
\tool{} flagged the keys (e.g., \texttt{/etc/ssh/ssh\_host\_rsa\_key}) as untracked because they are dynamically generated during package installation and do not appear in the package's manifest. 
An attacker could exploit this flaw to impersonate a legitimate build agent or execute man-in-the-middle attacks by intercepting or altering build instructions, injecting malicious code, or exfiltrating sensitive credentials and build artifacts. 
We coordinated disclosure with the Jenkins security team, who acknowledged and patched both vulnerabilities promptly, issuing two separate CVEs, each receiving a CVSS score of 9.1, along with an official security advisory.

\noindent \textbf{Embedded GitHub Token in acme.sh Image (CVE-2025-32111):}
The \texttt{acme.sh} Docker image (over 42,000 GitHub stars~\cite{acmesh}) contained a leaked GitHub token within its \texttt{.git/config} file. The token was introduced by a \texttt{COPY ./ .} instruction that unintentionally included the entire \texttt{.git} directory. Although a subsequent layer removed the file, the token persisted in Docker's layered filesystem and was retrievable from intermediate layers. \tool{} detected this transient but critical artifact through its deep-darkfiles analysis. The token granted unauthorized access to the project's GitHub resources, enabling an attacker to trigger CI workflows, tamper with source code, or modify release artifacts. Upon disclosure, the maintainers eliminated unnecessary file copies in their Dockerfile. The CVE received a CVSS score of 8.7.

These cases demonstrate that SDM surfaces concrete, exploitable vulnerabilities not merely theoretical gaps in widely deployed containerized environments.



\subsection{RQ3: Causal Origins}
Alongside the corpus-wide findings from RQ1 and the CVEs from RQ2, we identify three recurring origin categories for untracked files in containers.

\noindent \textbf{Ad-hoc materialization.}
Commands such as \texttt{curl | bash}, manual \texttt{COPY} operations, and unarchiving steps install software outside any package manager, producing binaries that never appear in an SBOM. 

\noindent \textbf{Build-context bleed-in and ephemeral retention.}
Broad operations like \texttt{COPY ./ .} or \texttt{rm -rf} on sensitive directories cause two related problems: they pull unintended files into the image, and they leave residual artifacts in intermediate layers even when a later instruction deletes them. 

\noindent \textbf{Post-build drift.}
Runtime initialization (e.g., first-run scripts that generate SSH keys, as in CVE-2025-32754/32755) and operational configuration changes introduce files after SBOM generation. 

\subsection{RQ4: Temporal Origins}
\label{sec:eval:origins}
To understand how Software Dark Matter evolves over time, we conducted a time series analysis of representative container image that consistently popular on DockerHub. Intuitively, most popular images are minimal, single-layer base images, such as Alpine or BusyBox. Instead, we illustrate our findings on the rare case of widely used more complex application that leverages a multi-layer build process and includes a variety of application-specific packages. In fact, it is based on Alpine and includes BusyBox - and with over 4.8 billion downloads it still ranks among the top 10 on DockerHub.

Figure~\ref{fig:grafana} shows that the proportion of Software Dark Matter can vary significantly between different versions. Our time series analysis considers a horizon looking back 5 years from today to when the project last changed its base image - coincidentally, when it switched to Alpine. In particular, the number of tracked files tends to increase over time as the size of packages increases. We observe a trend of files initially identified as dark in one version being gradually incorporated into packages in subsequent official releases. This suggests that the development and packaging of the project lag behind the content of the published container images.

Meanwhile, each new release introduces fresh untracked files, creating a continuous cycle in which the metadata never fully converges with the artifact. Packaging Lag demonstrates that even projects with active packaging efforts can exhibit persistent SDM simply because the upstream packaging cadence trails the image-publication cadence.




\begin{figure}[h]
\centering
\includegraphics[trim=10 40 10 10, width=.9\linewidth]{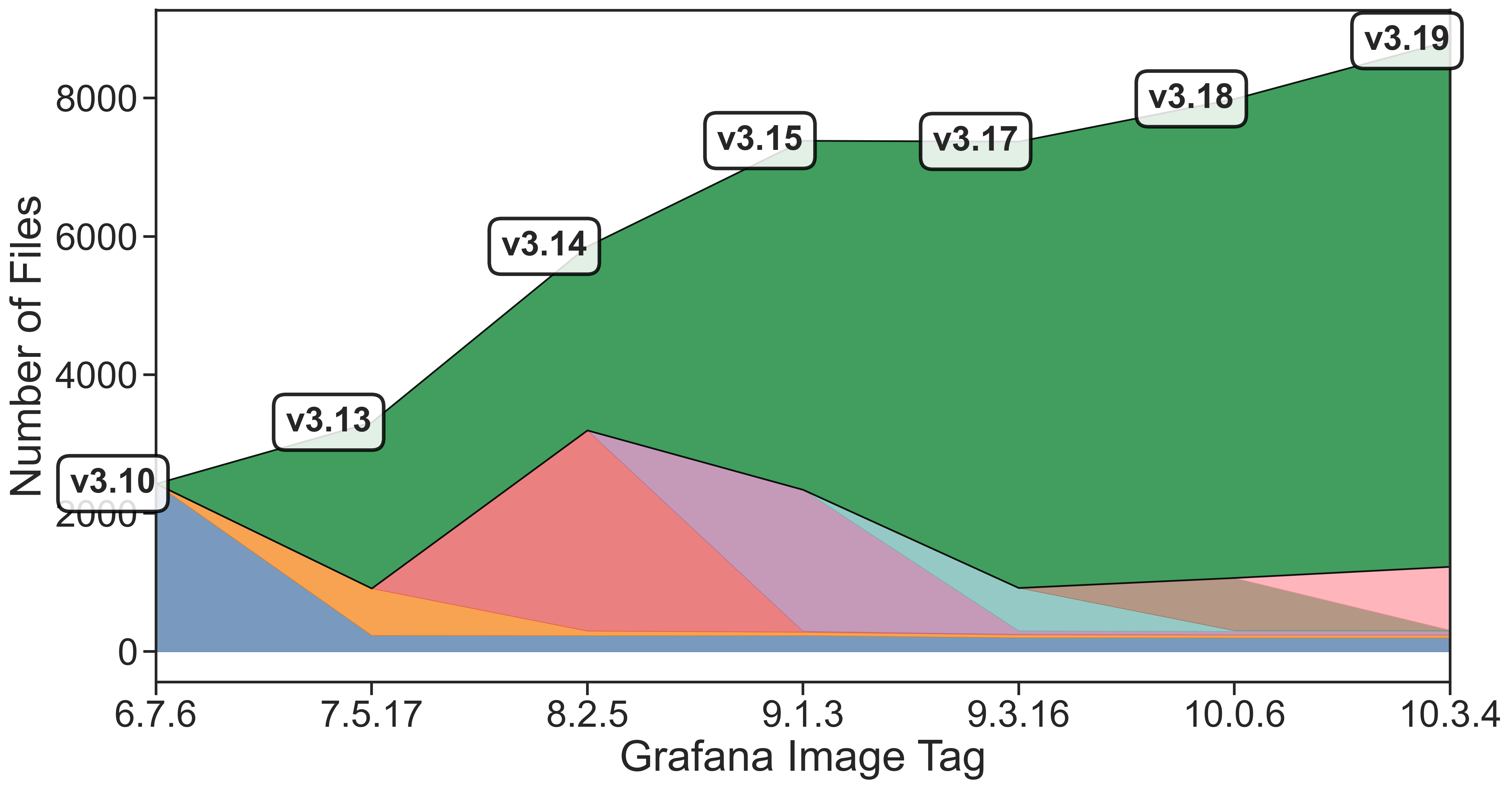}
\caption{Packaging lag of a representative container project - grafana/grafana over 3 years(2021-24)}
\label{fig:grafana}
\end{figure}

%% file: sections/maven.tex
Native language package ecosystems are one of the core building blocks of the aforementioned container images. There is an abundance of programming language specific package manager implementations, s.a. npm for JavaScript and pypi for Python. 
We choose to study the the popular Java dependency management system Maven. Java Maven is along lived ecosystems that focuses on enterprise applications and provides clear guidelines on meta-data release, e.g., POM files, semantic versioning and use of a centralized repository, Maven Central. It's maturity suggests absence of SDM and best in class SCA tool support, which makes it an ideal subject for our study.\cite{sotovalero_EMSE_2021, sotovalero_MSR_2019}

\paragraph{Data Collection and Methods}
Our evaluation confirms that Maven Central projects employ great dependency management conventions. We did not find any SDM in the top 1,000 Maven Central Projects. So, we refined our filter to collect all packages that indicate dependency bundling. Matching package names on a keyword list (e.g., -embedded-, -internal-, -thirdparty-, etc.) yield a dataset of 4,889 latest packages (e.g, 0.6\% of 803,000 Maven Central projects\cite{maven_central}). For each artifact we: (1) downloaded the JAR, source code, declared poml.xml (2) classify its files as: \emph{expected-external} if they matched a declared dependency’s package roots, \emph{self/metadata} if they belonged to the project itself, or \emph{untracked} if no provenance. This mirrors the core SDM definition used throughout the paper, measuring divergence between \emph{asserted} provenance and \emph{observed} artifact contents(Java bytecode and Maven metadata here). To better support Maven practices, we identify untracked dependencies based on nesting and shading indicators (See Section~\ref{sec:maven} on RQ3). 

\subsection{RQ1: Prevalence of Untracked Files}
SDM is prevalent in maven projects. We found indicators of SDM in 0.6\% of latest Maven Central packages. The actual prevalence of SDM follows a binomial distribution: Either a package does not hide dependencies, e.g., 80.4\% hide <10\%, or it hides most of its dependencies, e.g., 10.6\% hide >90\%. Treating each dependency equally, the mean number of dependencies is 11.91 and the mean number of untracked dependencies is 5.67. We do not uncompress untracked dependencies to files because the volume of files is meaningless when a whole package is untracked.
In our dataset, artifacts contain on average 11.91 dependencies, with roughly 5.67 hidden dependencies files per artifact. 
\begin{figure}[h]
    \centering
	\includegraphics[trim=0 40 0 20, width=1\linewidth]{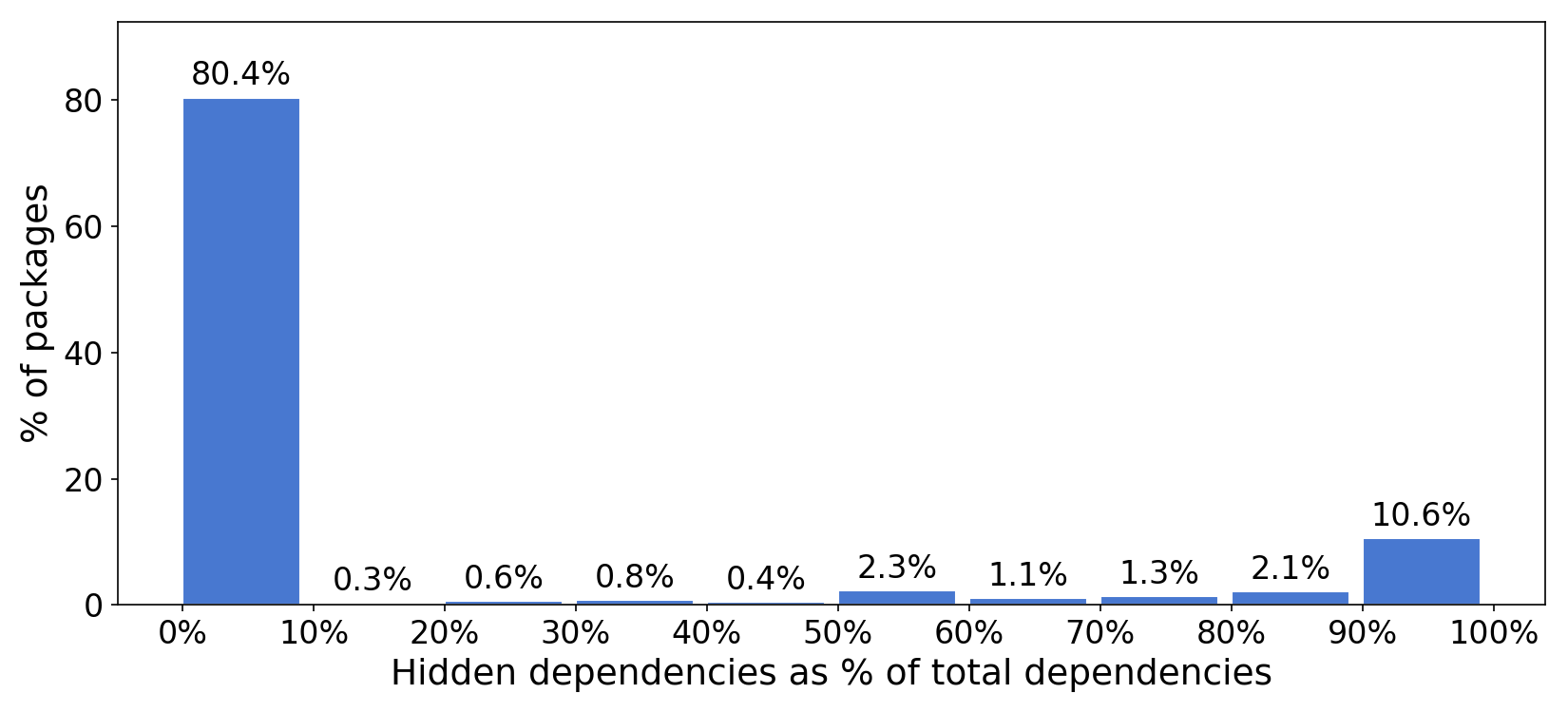}
	\caption{\% Untracked in suspicious maven projects}
	\label{fig:mavenprevalence}
\end{figure}

\subsection{RQ2: Security Impact} 
\begin{figure}[h]
    \centering
	\includegraphics[trim=0 50 0 40, width=1\linewidth]{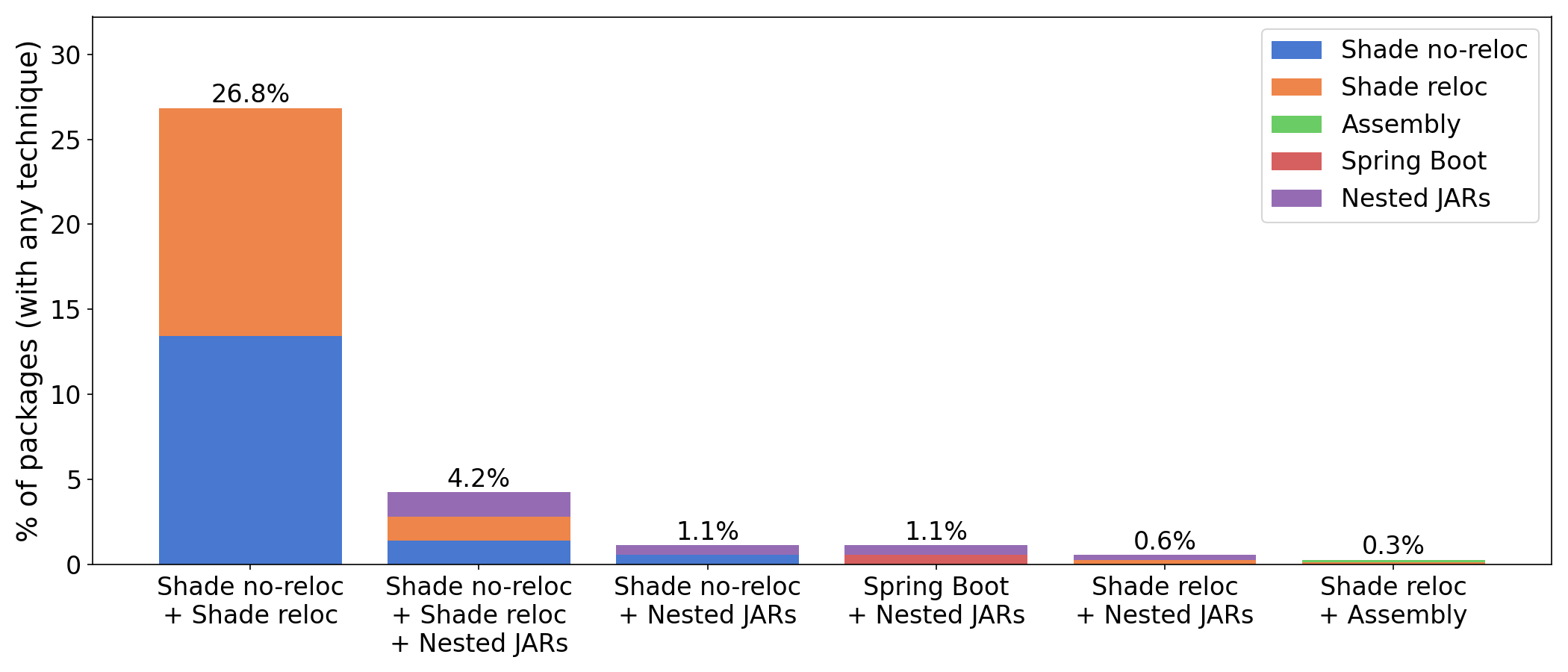}
	\caption{\% SDM in suspicious maven projects}
	\label{fig:mavenorigins}
\end{figure}

Maven is a popular ecosystem and shading and nesting are well-known parts of it. Hence not every untracked package is a security concern, and we can expect competitive SCA tools to implement custom heuristics to find shaded dependencies beyond analyzing the project meta-data. Figure~\ref{fig:mavenorigins} shows the distribution of CVEs found in the untracked dependencies. 
In absolute numbers, ~\tool{} identified 5,098 instances of untracked JARs at versions with known CVE associations.


\subsection{RQ3: Causal Origins} SDM in Maven projects is primarily driven by attempts to solve to version conflicts. We identify that projects either \emph{shade} or \emph{nest} their dependencies. 
In practice, shading is achieved by one of three methods: (1) using the maven-shade-plugin (2) using the maven-shade-plugin + renaming/ relocating (3) not using the maven-shade-plugin, but revealing shading by naming convention (e.g. using gradle toolchain and cross publishing to Maven Central).  Nesting is achieved bye one of three methods: (1) using the maven-assembly-plugin (2) using a framework like Spring Boot (3) Manually copied *.jar files in the file tree.

\subsection{RQ4: Temporal Origins}
Unlike containers, where SDM often arise from ad-hoc scripts or manual installs, Maven artifacts highlight a structural issue in the Java ecosystem: shading is intentional. SBOMs derived from POM metadata will systematically miss these embedded classes. Prior work has shown that "hidden" or outdated components can persist deep inside released artifacts for ecosystem- and architecture-specific reasons, including shading/ cloning and build-system indirections~\cite{dietrich2024, mir_SANER_23, pashchenko_ESEM_2018}. These studies often use fingerprinting or clone-style matching to rediscover vulnerable embedded components/classes post hoc. 


%% file: sections/extensions.tex
\label{sec:extensions}
We now turn to a class of artifacts absent from the prior analyses: \emph{Extensions}, also referred to as plugins or add-ons. Similar to to containers and native packages, extensions are distributed via registries, most commonly Marketplaces. This naming difference highlights the difference in intend between these ecosystems. Extensions require a host process/ platform they extend with non-standard features, and these non-standard features are commonly monetized. In conclusion, extensions are high stake and should be held to highest project management and security hygiene standards, making them another ideal target for our study. 
Recent studies have examined the behavioral security of CI plugins\cite{li_ccs_2024, gu_sp_2023} and IDE extensions\cite{lin_ndss_2024}, identifying over-privileged and vulnerable bundled artifacts. Our analysis is complementary, rather than studying what extensions do, we measure what they contain that is invisible to metadata-driven analysis. 
We study two instances: \emph{Jenkins plugins} (\texttt{.hpi} archives loaded into a CI/CD server JVM) and \emph{VS Code extensions} (\texttt{.vsix} archives loaded into a developer's editor via the Open~VSX registry~\cite{openvsx}). Both physically bundle third-party code whose identity is absent from any registry-facing manifest, making metadata-driven SCA underperform when analyzing these artifacts. 

The two platforms occupy opposite ends of the privilege spectrum and therefore pose different risks when SDM is present. 
Jenkins plugins execute server-side with access to build secrets, signing keys, and downstream deployment pipelines which means a single vulnerable bundled library can yield remote code execution on the build host. 
Jenkins also exposes a \emph{classloader inheritance} mechanism whereby Plugin~$A$ can load classes from Plugin~$B$'s bundled JARs at runtime, without any copy appearing in $A$'s own archive, an SDM propagation channel invisible to every existing SBOM model. 
VS Code extensions execute client-side on the developer's workstation, with access to the local file system, source code, and stored credentials. 
This can result in harm including credential theft, source-code exfiltration, and lateral movement into organizations. The same finding (a bundled dependency with a known vulnerability), therefore carries different exploitation paths depending on the platform.

\paragraph{Data Collection and Methods}~\\
\label{sec:extensions:data}
\noindent\textbf{Jenkins plugins.} We collected all {2066} plugins from the Jenkins Update Center; {1891} contained at least one nested JAR. \tool{}, configured for the nesting and shading heuristics (Section~\ref{sec:maven}) recovered {1501}~JARs across {425}~plugins that lacked standard metadata.

\noindent\textbf{VS Code extensions.} We collected metadata for the top \~{3,000} extensions on Open~VSX prioritized by download count, downloading and scanning each \texttt{.vsix} artifact. Each extension's declared dependency set was extracted from its root \texttt{package.json}. To identify undeclared bundled packages, we used four complementary signals: shipped \texttt{node\_modules/} subtrees (which retain per-package \texttt{package.json} files giving exact identity), webpack \texttt{*.LICENSE.txt} sidecars (from which version strings can be speculatively extracted for ${\sim}$10\% of bundled packages) and and source-map paths in \~570 extensions.  Root-level lock files were present in \~100 extensions but contributed little additional signal, as most appeared to be stale artifacts from upstream repositories rather than authoritative records of the bundled content.

\subsection{RQ1: Prevalence of Untracked Files}
\label{sec:extensions:rq1}

\noindent\textbf{Jenkins plugins.} After full transitive dependency expansion, the mean untracked percentage across {1102}~plugins is \textbf{10.0\%} (median 0\%). A total of {1737}~untracked JAR instances remain across {318}~plugins, and {74}~plugins have a majority of their bundled JARs untracked. An additional {57}~plugins bundle a JAR at a version that differs from what their POM declares making the SBOM not merely incomplete but actively misleading.

\noindent\textbf{VS Code extensions.} Among the 1,249~analyzed extensions, the majority that ship dependencies do so without declaring them: fewer than 2\% populate the npm \texttt{bundledDependencies} field, and the VSIX manifest carries no structured record of bundled code. Critically, we can only positively assert bundled file provenance for the subset of extensions that ship raw \texttt{node\_modules/} directories, source maps with pnpm-style versioned paths(source maps from extensions built with pnpm embed the package version directly in the content-addressed store path), or webpack license comments—a small fraction of the ecosystem. For these, every undeclared package is individually identifiable yet invisible to any metadata-driven SCA tool. The majority of extensions, however, use a JavaScript bundler (\texttt{webpack}, \texttt{esbuild}, \texttt{rollup}) that concatenates all dependencies into a single minified output, destroying per-package attribution entirely. Recovering provenance from these artifacts would require "unbundling" by segmenting the minified bundle into per-module spans and matching each against fingerprints (e.g., locality-sensitive hashes such as TLSH~\cite{tlsh}) computed over the full npm package corpus which is a substantially harder problem than Maven shading, where \texttt{.class} files retain their original package namespaces, and one we scope as future work. Our prevalence figures for openvsx is therefore a conservative lower bound restricted to the observable fraction of the registry.


\subsection{RQ2: Security Impact}
\label{sec:extensions:rq2}

\textbf{Jenkins plugins.} Cross-referencing all resolved JARs against a database of {20}~widely-exploited library families identified 648~vulnerable JAR instances across 381~plugins, covering approximately 200,000~installations. Of these, 54.6\% are invisible or misleading to SBOM-based scanners: either entirely absent from declared dependency metadata or declared at a safe version while the physically bundled JAR is vulnerable. The most affected libraries include Apache HttpClient (189~plugins), \texttt{jackson-databind} (140~plugins), and SnakeYAML (31~plugins, CVSS~Critical).

Beyond direct bundling, the classloader inheritance graph exposes 80~additional vulnerable JAR instances affecting 48~plugins whose archives contain no copy of the vulnerable JAR. Among the 33~plugins uniquely surfaced through this channel are high-installation maintained plugins such as \texttt{sonar} (50,476~installs) and \texttt{pipeline-maven} (26,768~installs). Notably, plugins marked adopt-this-plugin or deprecated which account for 33 of 84~CRITICAL instances despite serving only ${\sim}$12\% of the install base remain depended upon by actively maintained downstream plugins, propagating their untracked vulnerabilities across the plugin graph with no active maintainer to remediate.

\textbf{VS Code extensions.} Among 187~extensions with identifiable untracked packages, we found {1259} packages matching known CVEs (132~CRITICAL, 619~HIGH, 508~MODERATE/LOW), all absent from the extension's declared dependencies. 
We identified a top-downloaded extension [redacted] that ships 88 untracked packages against only 14 declared, including dependencies at versions affected by high-severity CVEs such as prototype pollution and denial-of-service vulnerabilities. The extension's architecture suggests a plausible attack surface through which these untracked vulnerabilities could be exploited. We are currently in coordinated disclosure with the extension's maintainers and withhold identifying details until remediation is complete.

Beyond vulnerable packages, our file-level enumeration surfaces build-context bleed-in invisible to advisory-driven scanning: extensions shipping \texttt{.claude/} and \texttt{.copilot/} directories (AI-assistant session data and cached credentials), \texttt{.env} files, \texttt{.ssh/} directories, and residual \texttt{.git/} histories, a parallel to CVE-2025-32111 (Section~\ref{sec:container}), now distributed through a trusted extension marketplace.

\textbf{Responsible Disclosure.} We have reported all vulnerabilities identified to the affected parties. For Jenkins plugins, disclosures were coordinated with the Jenkins security team, including plugins maintained by the Jenkins project itself. For VS Code extensions, we contacted the individual extension maintainers. Disclosures are currently under review and patches are expected in upcoming releases. At the time of writing, disclosures remain under review and identifying details are withheld pending remediation. If accepted we will update this section with full details.


\subsection{RQ3: Causal Origins}
\label{sec:extensions:rq3}

\textbf{Jenkins plugins.} Untracked JARs in Jenkins originate primarily from pre-Maven-era artifacts (which lack standard coordinates and are unresolvable by any declarative SBOM tool), split-modular Maven projects whose sub-module JARs are not enumerated by the parent POM, and JARs placed directly into \texttt{WEB-INF/lib/} to pin a version or incorporate a vendor SDK outside Maven's resolution. 
The common thread is that all three paths bypass Maven's dependency model, so build-time SBOM tools which consume the Maven model as their source of truth never see these files.

\textbf{VS Code extensions.} SDM in VS Code extensions compounds from four failures: (1)~the \texttt{.vsixignore} exclusion mechanism is opt-in and widely omitted, sweeping developer credentials and tooling state into the artifact; (2)~the \texttt{bundledDependencies} field in \texttt{package.json} (npm mechanism explicitly designed for it) is populated by fewer than 2\% of extensions; (3)~the VSIX manifest records no dependency information; and (4)~JavaScript bundlers erase per-package attribution, requiring the corpus-scale unbundling approach described in RQ1 to recover provenance. 


\subsection{RQ4: Temporal Origins} 
\label{sec:extensions:rq3}
\begin{figure}[h]
    \centering
	\includegraphics[trim=0 40 0 40, width=\linewidth]{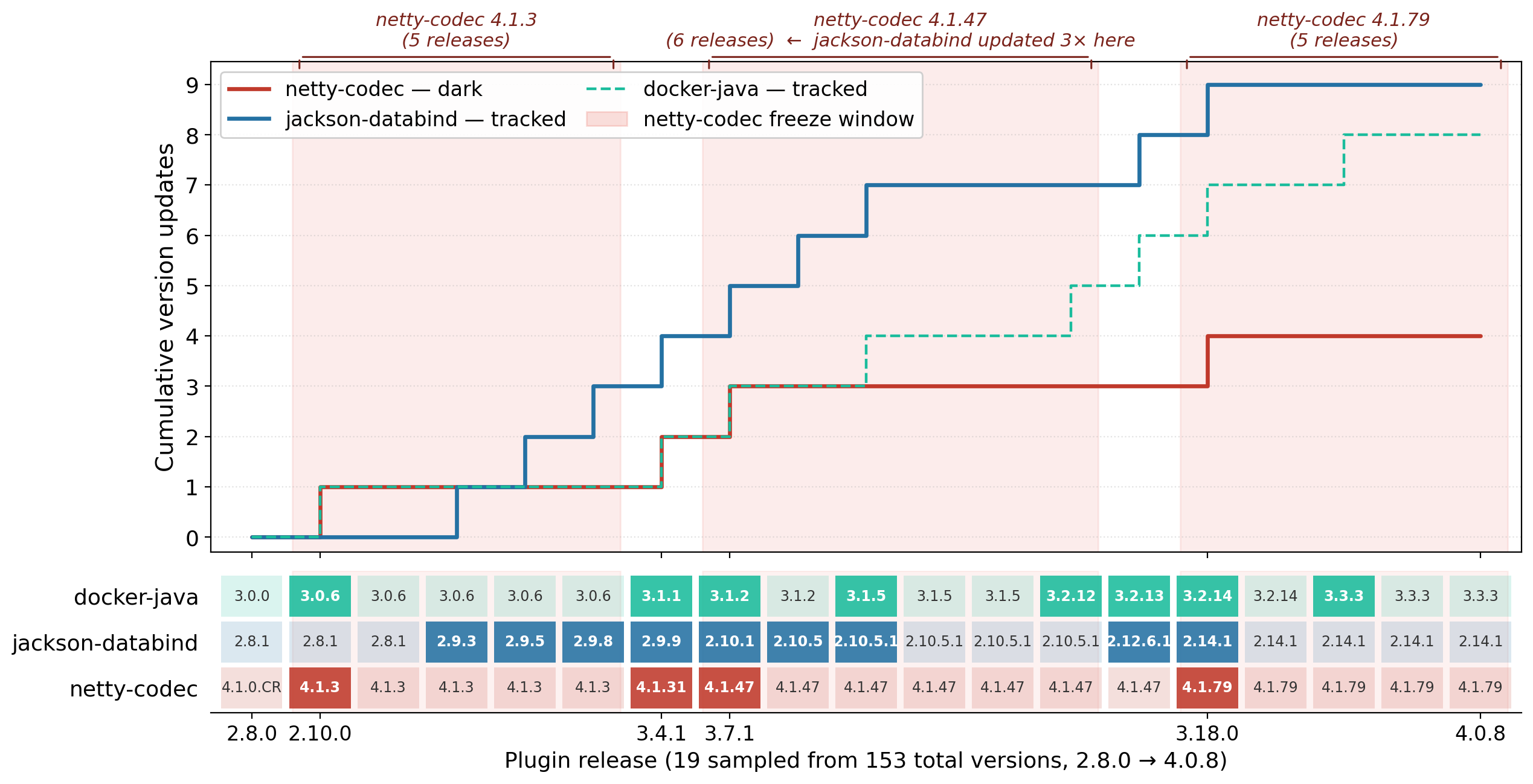}
	\caption{Jenkins plugin: update cadence of dark vs. tracked JARs across versions}
	\label{fig:jenkins-update-cadence}
\end{figure}

\textbf{Jenkins plugins.} 
Consistent with the Maven analysis, SDM in Jenkins plugins does not disappear over time. Figure~\ref{fig:jenkins-update-cadence} illustrates this through a plugin(most dark jars amongst the most popular), Jenkins Artifactory with 153 versions from 2.8.0 to 4.0.8. 
It has tracked dependencies such as \texttt{docker-java} and \texttt{jackson-databind} receive frequent version bumps, while untracked ones such as \texttt{netty-codec} are updated measurably less often. However, the untracked dependencies \emph{are} updated, which implies that plugin maintainers are aware of their existence and perform at least some dependency hygiene on them. This reinforces the Maven finding that SDM is not accidental but systemic: developers knowingly embed and maintain these dependencies, yet the artifacts they produce offer no metadata through which downstream consumers could discover, audit, or respond to vulnerabilities in them. 
The producer has visibility but the consumer is blind and no amount of SBOM tooling applied at the consumer side can close that gap without cooperation from the producer's build pipeline.


\textbf{VS Code extensions.} For causal inference, we used source maps as a sidechannel that provides sufficient information, but we were not able to conduct a longitudinal analysis of VS Code extensions on this data source.

%% file: sections/enterprise.tex
The preceding sections measured SDM in public container images, Maven artifacts, and extension marketplaces. To test whether the phenomenon generalizes to fully managed infrastructure, we apply \textsc{Darkfiles} to production hosts at a large software company (\textasciitilde9{,}000 engineers) that operates organization-wide SCA integration, persisted CI metadata, and deploys on a commercial Linux distribution under a maintenance agreement. In this setting where every file on managed hosts is expected to be accounted for, SDM serves as a proxy for provenance-coverage gaps in the deployment pipeline.

We adapt the \textsc{Darkfiles} workflow as follows. The asserted state is derived by aggregating OS-level package metadata and stored CI-generated SBOMs. Proprietary application bundles installed in locked-down directories via the managed pipeline are trusted and excluded. The observed state is inventoried at fleet scale from the enterprise EDR file tables~\cite{} (paths, types, permissions, timestamps, hashes), replacing \textsc{Darkfiles}'s own crawler. We apply the standard exclusions from Section~5 (transient paths such as \texttt{/dev}, \texttt{/run}, caches, logs, and UI assets), normalize paths and symlinks, align asserted and observed views, and compute per-host diffs of untracked, missing, and hash-mismatched files.


After de-duplicating untracked files across hosts, we traced their provenance using the company's internal VCS and configuration repositories. Infrastructure-as-Code (IaC) and configuration-management frameworks are designed to declaratively specify system state, translating high-level infrastructure requirements into reproducible deployments. However, our analysis revealed that these systems are the dominant source of SDM inline with Rahma et al work \cite{rahman_ICSE_2019, rahman_TOSEM_2021, rahman_TOSEM_2023}. Provisioning runs routinely retrieve and place external artifacts(binaries, scripts, and configuration fragments) from remote hosts outside the package manager's purview, producing files whose provenance is captured by no metadata channel. Most notably, we detected downloads originating from abandoned hosting infrastructure that had subsequently been taken down, rendering the provenance trail for those files entirely unverifiable. SDM further included operational tooling for telemetry and monitoring: scripts critical to fleet operations yet invisible to the SBOM-driven governance pipeline.

These findings confirm that SDM persists even under best-practice enterprise conditions (RQ1) and that its primary origin in managed environments is post-deployment drift like provisioning workflows, IaC-driven fetches, and out-of-band operational changes that introduce files after build-time SBOM generation (RQ3). The security implication (RQ2) is that SBOM-driven governance cannot assume completeness at build time. It requires a continuous reconciliation loop to detect drift in long-lived infrastructure. Following our report, the company responded by packaging previously untracked files, eliminating unnecessary bloat, and integrating \textsc{Darkfiles}-style validation into their provisioning pipeline, demonstrating that SDM analysis yields actionable remediation even in mature, centrally managed environments.

%% file: sections/discussion2.tex

Software Dark Matter is ultimately a consequence of the informational gap between what a build pipeline produces and what its metadata captures.
Across the ecosystems we studied, SDM consistently traces back to a small number of pipeline stages -- a careless copy instruction, a shading plugin, a provisioning script that fetches binaries out-of-band -- yet these stages are precisely the ones invisible to metadata-centric tooling.
Once surfaced, each instance of dark matter presents maintainers with a concrete choice: fold it into the SBOM, remove it, or accept it as a known gap.
The key insight is that this triage does not require new tooling per artifact type.
The same provenance-and-reachability workflow generalizes from container layers to shaded JARs to bundled extensions to enterprise hosts.
A ``\tool{}-before-SBOM'' step in the build pipeline converts dark matter from an undisclosed omission into an explicit quality signal: either the provenance gap is resolved before the SBOM is published, or it is labeled as an open issue that warrants attention.
Over time, this feedback loop also improves tooling itself, as engineers can identify recurring patterns of metadata loss and build targeted fixes into their pipelines and generators.

Addressing SDM instance by instance, however, is insufficient.
The root cause is that current SBOM generation is decoupled from supply chain processes: tools reconstruct provenance after the fact from metadata that was never designed to be complete.
Eliminating dark matter systematically requires tightly coupling transparency metadata generation to the build itself, so that provenance is captured at the point of creation rather than inferred later.
Efforts like SBOMit~\cite{sbomit} point in this direction by leveraging higher-fidelity build attestations, such as those produced by in-toto~\cite{torres_2019} and its Witness~\cite{witness} implementation, to produce SBOMs grounded in what the build actually did.\cite{newman_sigstore_2022, amusuo_icse_2025}
Similarly, Debian's \texttt{buildinfo} ties package contents to a reproducible build record.
Contemporary work like LastPyMile~\cite{lastpymile}, PyRadar~\cite{pyradar}, and CovSBOM~\cite{zhao2024covsbom} has independently identified step-specific instances of divergence between artifact contents and declared metadata.
The SDM framing unifies workstream under a single theoretical framework, and that \tool{} provides the measurement infrastructure to evaluate whether these emerging approaches succeed in closing the gap.

%% file: sections/conclusion.tex
In this paper, we introduced Software Dark Matter (SDM), a general-purpose metric that quantifies the divergence between the files physically present in a software artifact and those asserted by its metadata. 
We measured SDM using \tool{}, a modular analysis tool that combines exhaustive file extraction, a large-scale file-to-package index, and a static reachability filter to surface the security-relevant subset of untracked files.
We evaluated \tool{} across disjoint ecosystems - DockerHub containers, Maven Central artifacts, Jenkins plugins, VS Code extensions, and a production enterprise environment -- and demonstrated that SDM is not an edge case but a prevalent, structurally rooted phenomenon arising from recurring patterns: build-context bleed-in, provenance loss through repackaging, metadata divergence, and post-build drift. 
We observed "Packaging Lag", in which metadata trails artifact contents across multiple releases before converging. 
The security implications are concrete: SDM harbors CVEs invisible to SBOM-driven pipelines, and our analysis led to the zero-day discovery of three confirmed high-severity vulnerabilities in widely deployed projects. 
These findings show that incomplete SBOMs do reduce transparency and cause a false sense of security on deployed software.
SBOMs remain the most promising mechanism for supply chain transparency, but the tools that populate them rely on metadata that was never designed to be complete. 
Addressing SDM requires both filesystem-grounded validation integrated into build pipelines and a rethinking of development practices to minimize divergence between what is built and what is declared. 
We release \tool{} as open-source infrastructure to support this effort.